\documentclass[10pt,final,twocolumn,journal]{IEEEtran}
\usepackage{filecontents,lipsum}
\usepackage[noadjust]{cite}
\usepackage[percent]{overpic}
\usepackage{stfloats}
\usepackage{ctable}
\usepackage[mathlines]{lineno}

\usepackage{cite}
\usepackage{graphicx}
\usepackage{caption}
\usepackage{subcaption}
\usepackage{authblk}
\usepackage{xcolor}
\usepackage{amsmath}
\usepackage{array}
\usepackage[normalem]{ulem}

\newcommand*{\Scale}[2][4]{\scalebox{#1}{$#2$}}%
\begin{document}
\title{Dielectric Engineered Tunnel Field-Effect Transistor}
\author{Hesameddin Ilatikhameneh, Tarek A. Ameen, Gerhard Klimeck, Joerg Appenzeller, and Rajib Rahman
\thanks{This work was supported in part by the Center for Low Energy Systems Technology (LEAST), one of six centers of STARnet, a Semiconductor Research Corporation program sponsored by MARCO and DARPA.}
\thanks{The authors are with Purdue University, West Lafayette, IN, 47907 USA e-mail: hesam.ilati2@gmail.com.}
}
%



\maketitle
\setlength{\textfloatsep}{7pt plus 1.0pt minus 2.0pt}
\begin{abstract}

The dielectric engineered tunnel field-effect transistor (DE-TFET) as a high performance steep transistor is proposed. In this device, a combination of high-k and low-k dielectrics results in a high electric field at the tunnel junction. As a result a record ON-current of about 1000 uA/um and a subthreshold swing (SS) below 20mV/dec are predicted for WTe$_2$ DE-TFET. The proposed TFET works based on a homojunction channel and electrically doped contacts both of which are immune to interface states, dopant fluctuations, and dopant states in the band gap which typically deteriorate the OFF-state performance and SS in conventional TFETs. 
\end{abstract}

\section{Introduction}
Tunnel FETs (TFETs) are strong candidates for future electronics applications due to their promise of low power consumption originating from their subthreshold-swing (SS) being less than the conventional limit of 60 mV/dec \cite{Appenzeller1}. However, TFETs usually suffer from lower ON-currents if compared to ultra-scaled MOSFETs \cite{Sub12}, since their currents are the result of the tunneling process. To overcome the low ON-current challenge, two obvious solutions are: 1) Increasing the electric field at the tunnel junction, 2) decreasing the band gap and effective mass of the channel material. However, reducing the band gap increases the OFF-current of the device and limits the maximum allowable supply voltage V$_{\rm DD}$. Accordingly, there is a lower limit on the band gap for any V$_{\rm DD}$ ($E_g \geq qV_{\rm DD}$).

Previous approaches for increasing the electric field at the tunnel junction included using an internal field (i.e. piezo-electric field) in nitride heterostructures \cite{Wenjun} or employing 2D channel materials with tight gate control \cite{Hesam1, Fiori}. In this work, a new method for increasing the electric field is proposed. It is shown that using a low-k dielectric next to high-k dielectrics can, in principle, increase the electric field in homo-junction TFETs significantly. The structure of the dielectric engineered TFET (DE-TFET) is shown in Fig. 1a. At the interface between the dielectrics, the displacement vector should be continuous ($\epsilon_1 E_1 = \epsilon_2 E_2$). 
With $\epsilon_2 \ll \epsilon_1$, the low-k dielectric has a larger electric field $E_2$ 
as illustrated in Fig. 1b. Since the high electric field region occurs right at the top of the tunnel junction, the ON-current of the device is amplified. Notice that the same idea (combination of high-k and low-k dielectrics) can be used in other device structures; e.g. gate-all-around, double and single gated TFETs. 
 
\begin{figure}[!t]
        \centering
        \begin{subfigure}[b]{0.25\textwidth}
               \includegraphics[width=\textwidth]{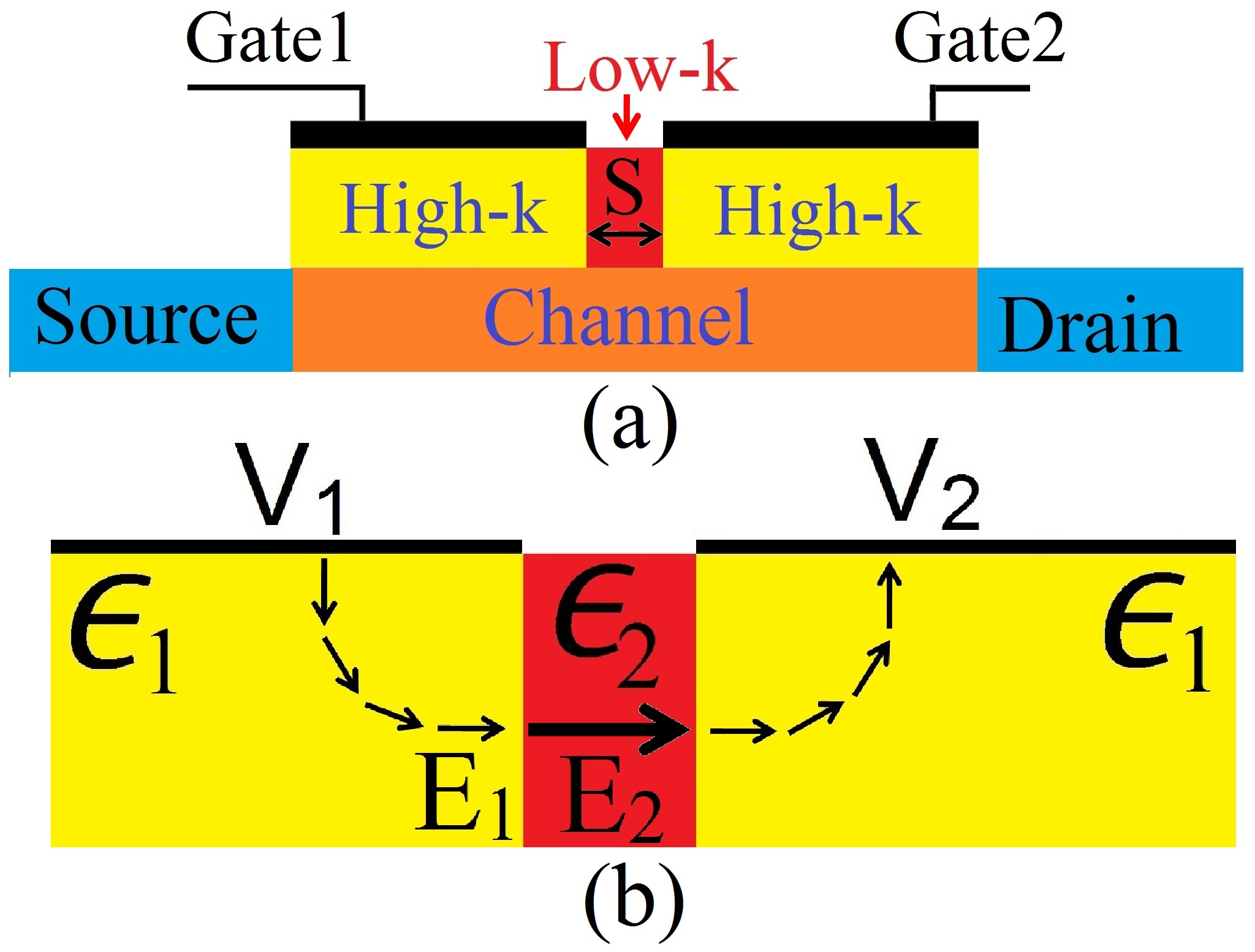} 
                \label{fig:struct1}
        \end{subfigure}%
        \begin{subfigure}[b]{0.25\textwidth}
               \includegraphics[width=\textwidth]{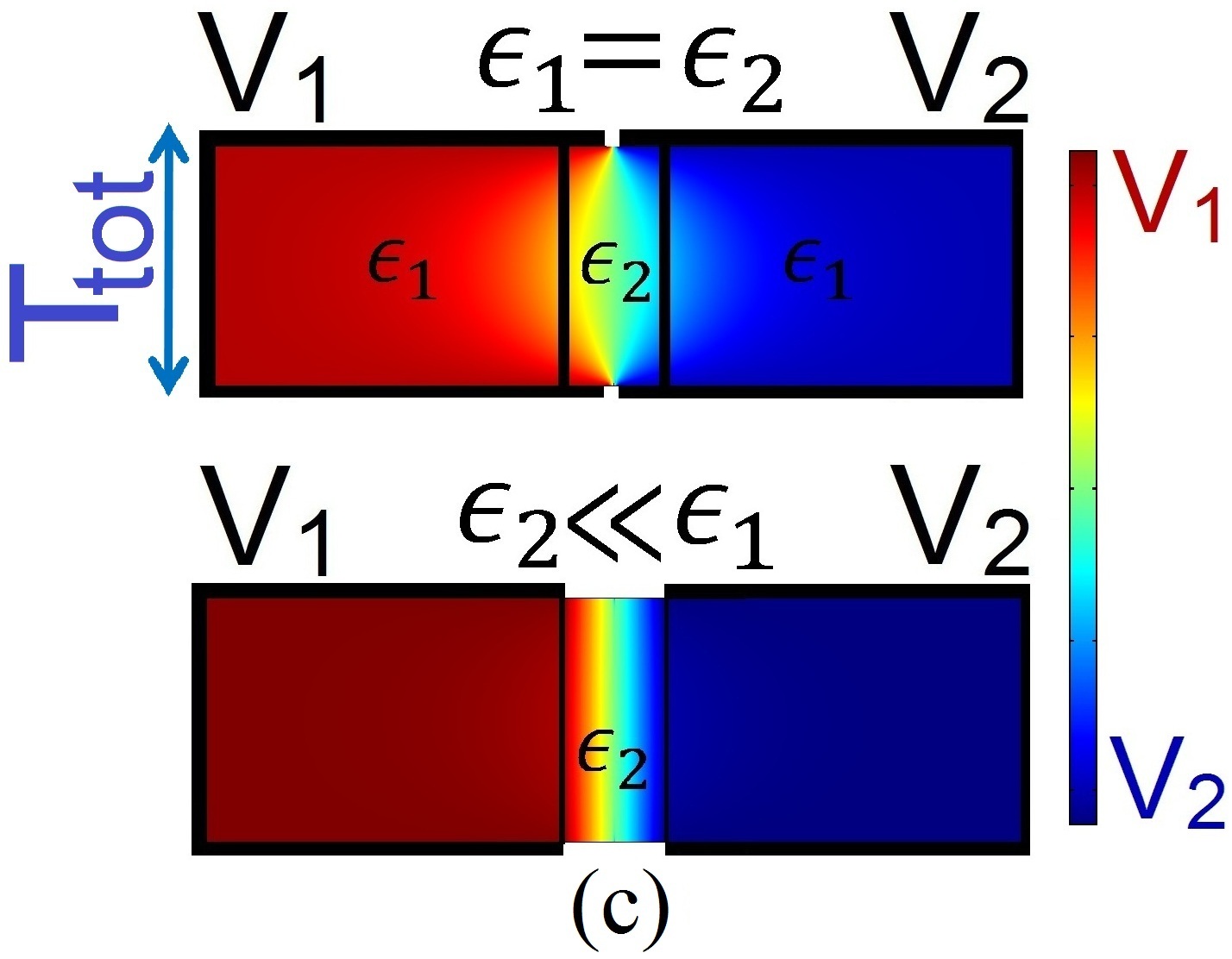}
                \label{fig:Same_EoT}
        \end{subfigure}%
        \caption{a) Physical structure of a DE-TFET. b) Illustration of electric field amplification using a low-k dielectric next to high-k dielectrics. c) Potential profile in a conventional ED-TFET versus DE-TFET.}\label{fig:Fig1}
\end{figure}

\begin{figure}[h!]
        \centering    
        \begin{subfigure}[b]{0.25\textwidth}
               \includegraphics[width=\textwidth]{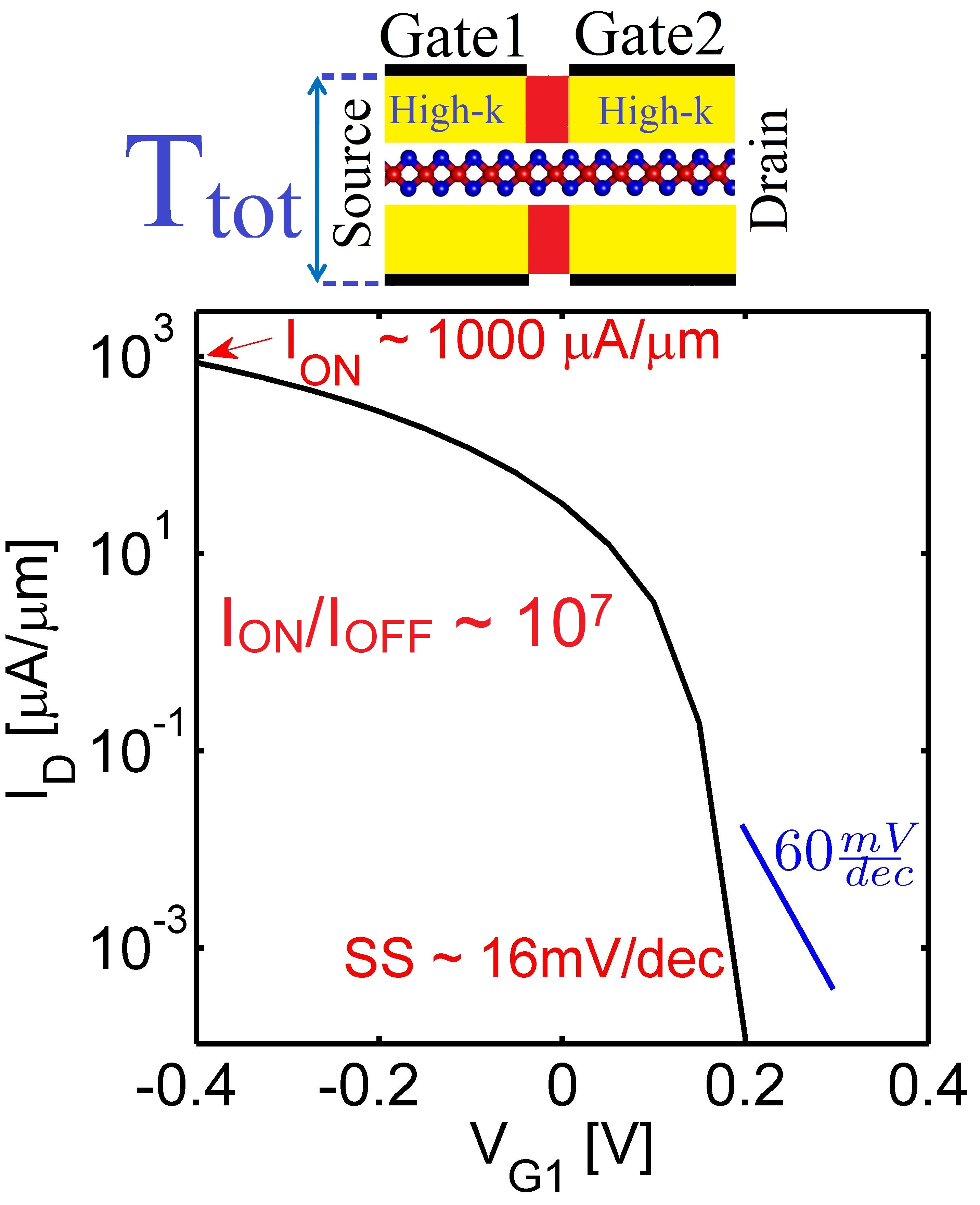}
               \vspace{-1.5\baselineskip}
                \caption{}
                \label{fig:Same_EoT}
        \end{subfigure}%
        \begin{subfigure}[b]{0.24\textwidth}
               \includegraphics[width=\textwidth]{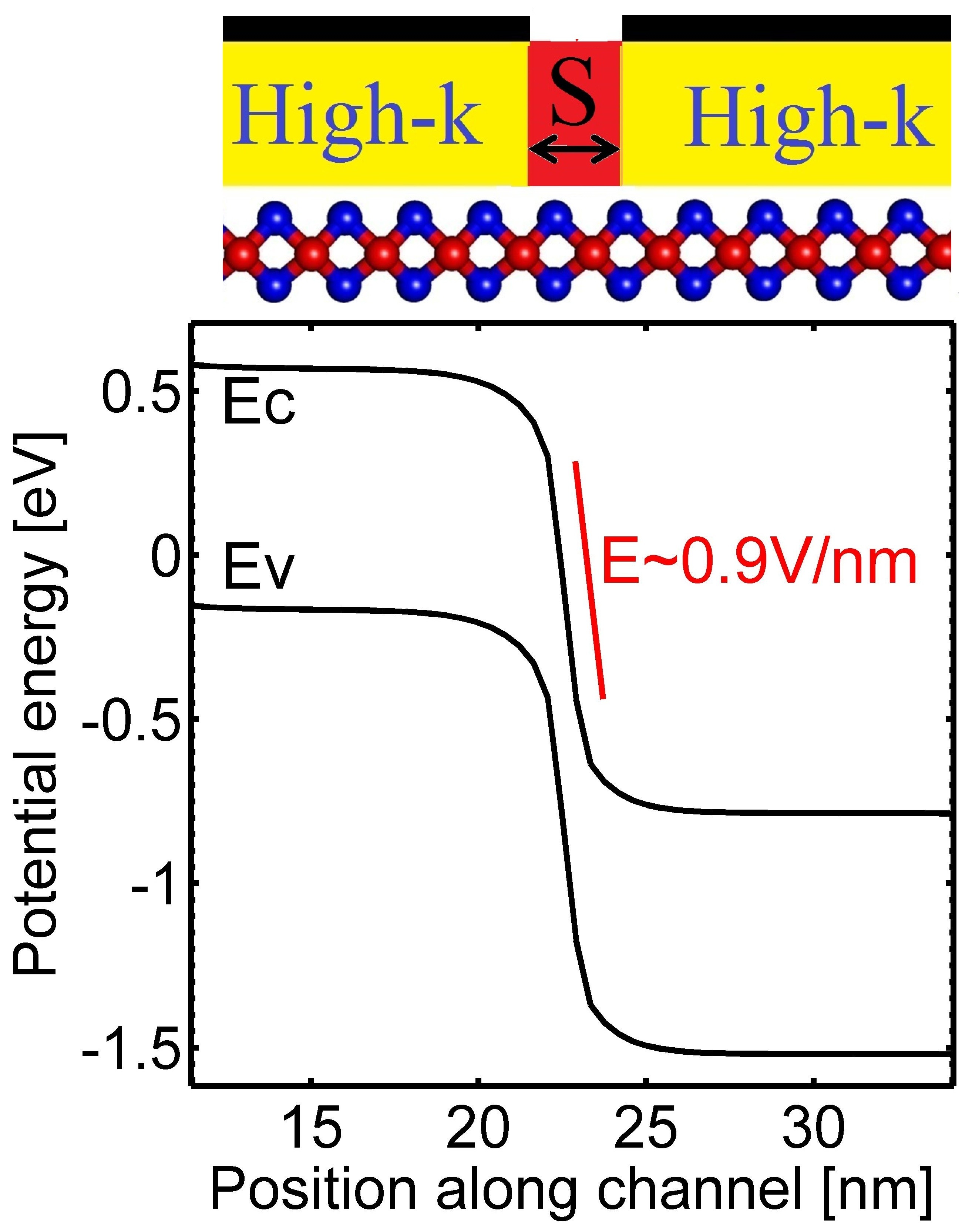}
               \vspace{-1.5\baselineskip}
                \caption{}
                \label{fig:Spacing}
        \end{subfigure}%
        \quad            
        \begin{subfigure}[b]{0.25\textwidth}
               \includegraphics[width=\textwidth]{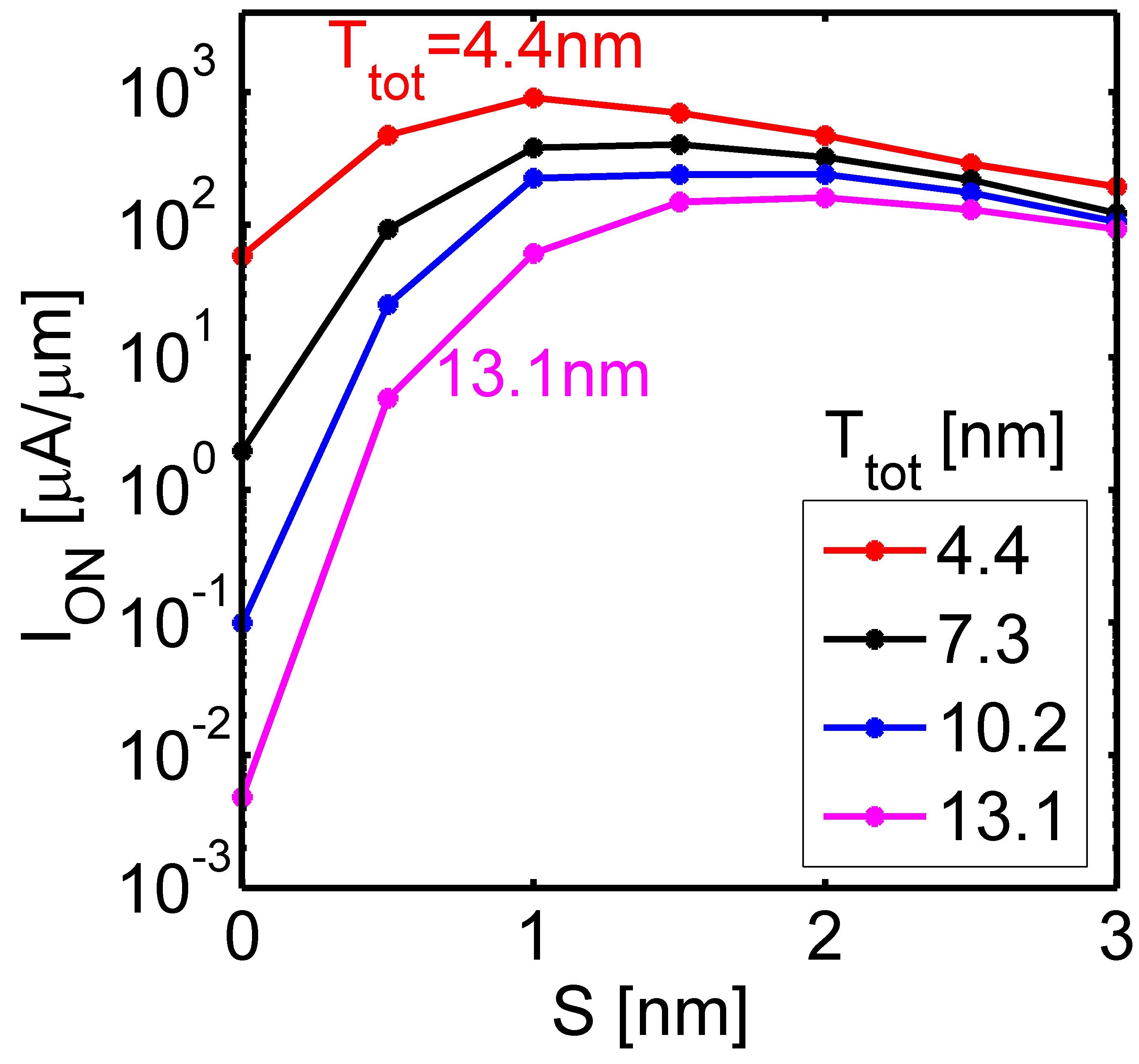} 
               \vspace{-1.4\baselineskip}
                \caption{}
                \label{fig:struct1}
        \end{subfigure}%
        \begin{subfigure}[b]{0.25\textwidth}
               \includegraphics[width=\textwidth]{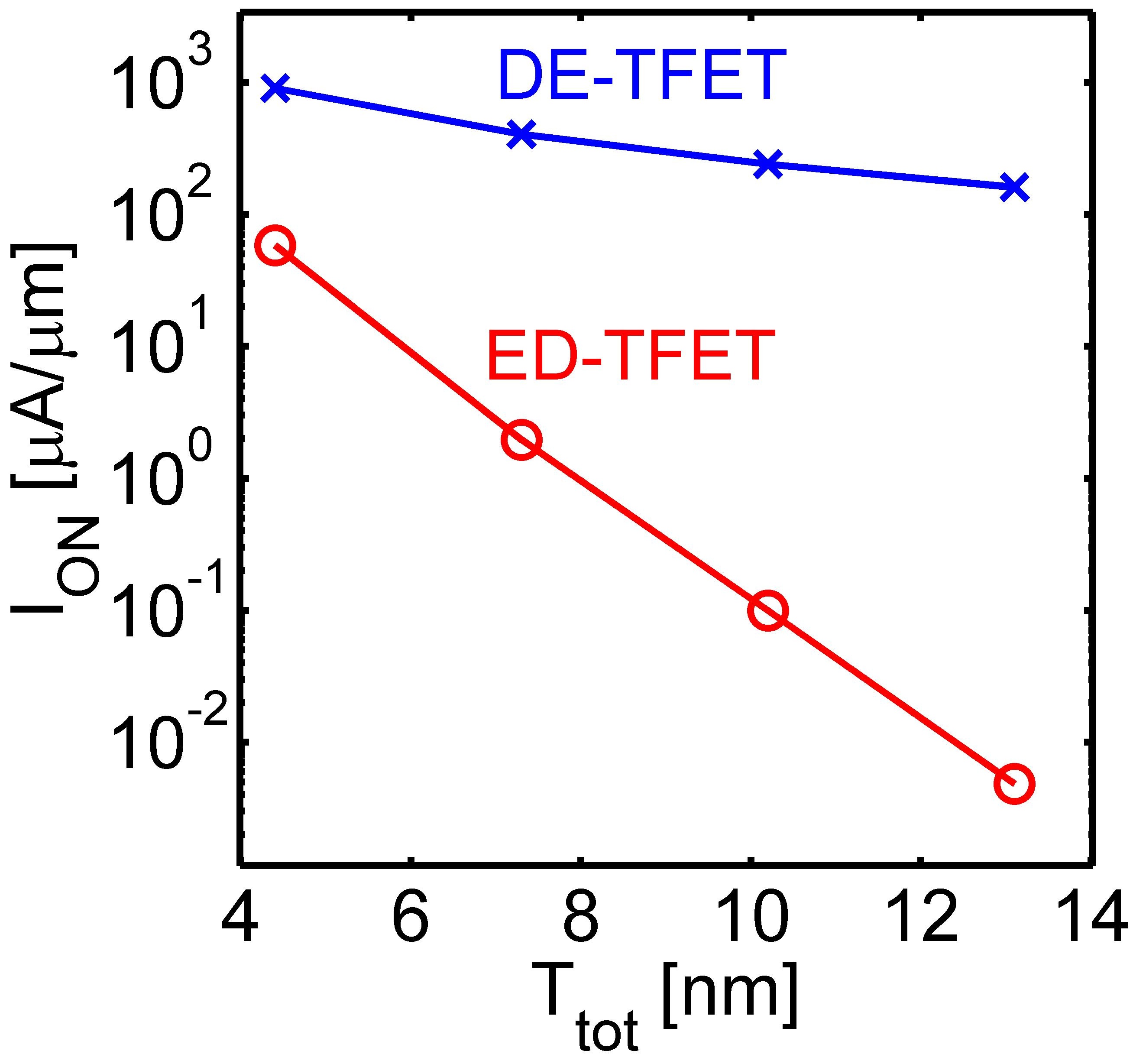} %
               \vspace{-1.4\baselineskip}
                \caption{}
                \label{fig:struct1}
        \end{subfigure}%
        
        \vspace{-0.3\baselineskip}
        \caption{a) Physical structure, transfer characteristics, and b) band diagrams of a mono-layer WTe$_2$ double gated DE-TFET with left and right oxide lengths of 12nm. $\rm V_{G2}$ is fixed to 1V. c) The dependence of the ON-current of a WTe$_2$ DE-TFET on the spacing for different $T_{tot}$. d) Comparison of performance sensitivity to $T_{tot}$ of WTe$_2$ DE-TFET with ED-TFET.}\label{fig:Fig2}
\end{figure}

Fig. \ref{fig:Fig1}c compares the potential profile for a \emph{conventional} electrically doped TFET (ED-TFET) \cite{Hesam2} with that of a DE-TFET. In case of the ED-TFETs, the width of the potential spread is proportional to the total thickness of the device $T_{tot}$ (the body thickness plus the oxide thickness shown as $T_{tot}$ in Figs.  \ref{fig:Fig1}c and \ref{fig:Fig2}a). In contrast, most of the potential drop in the DE-TFET occurs over the width of the low-k dielectric region (labeled as $S$ in Fig. \ref{fig:Fig1}a). Equation (\ref{eq:opt1}) compares the average electric field of a \emph{conventional} ED-TFET with that of the DE-TFET showing a possible gain in the electric field for the DE-TFET (as long as $S $ $<$ $4T_{tot}/\pi$).
\begin{equation}
\label{eq:opt1}
\langle E_{DE-TFET} \rangle \approx \frac{V_1-V_2}{S} > \langle E_{ED-TFET}  \rangle \approx \frac{\pi}{4}\frac{V_1-V_2}{T_{tot}}
\end{equation}

Furthermore, the DE-TFET offers several other advantages: 1) The performance is not very sensitive to the spacing (shown as $S$ in Fig. 1) when $S$ is chosen around the optimum value, in contrast to the \emph{conventional} ED-TFET in which the ON-current decreases monotonically with increasing $S$ \cite{Hesam2} (i.e. lack of optimum $S$: $\frac{dI_{ON}}{dS} \neq 0$). 2) Increasing the oxide thickness does not deteriorate the performance of the device, while the performance of electrically doped devices strongly depends on the thickness \cite{Hesam2}. 3) There is no need for chemical doping which means the DE-TFET does not exhibit the drawbacks of chemically doped devices \cite{sapan} (i.e. dopant states in the bandgap which deteriorate the OFF-state performance and SS in addition to the disorder introduced by fluctuations in dopant locations). Note that despite the oxide thickness having less impact on the performance of the DE-TFETs compared to ED-TFETs, a thinner oxide is still slightly beneficial for an improved ON-state. 
 


\setlength{\belowdisplayskip}{4pt} \setlength{\belowdisplayshortskip}{4pt}
\setlength{\abovedisplayskip}{4pt} \setlength{\abovedisplayshortskip}{4pt}


In Section III, full-band quantum transport simulations are performed on a monolayer WTe$_2$ homojunction DE-TFET to investigate its performance. The performance sensitivity of the DE-TFETs to the spacing, thickness, and channel material (WSe$_2$ and WTe$_2$) is discussed by both analytic modeling and numerical simulations in Section IV.  
\section{Simulation details}
To represent WSe$_2$ and WTe$_2$ as channel materials of the DE-TFET atomistically, we have utilized a sp$^3$d$^5$ 2nd nearest neighbor tight-binding model with spin-orbit coupling. Atomistic quantum transport simulations employing a self-consistent Poisson-NEGF (Non-equilibirum Green's Function) method have been utilized. The details of the simulation methods and material properties can be found in \cite{Hesam1}. The simulated double gated DE-TFETs assume a structure shown in Fig. \ref{fig:Fig1}a. Each gate has a length of 12nm, with the low-k spacing $S$. Notice that the low-k dielectric fills the space between the gates, right on top of the tunnel junction. A source-drain voltage $V_{DS}$ of 0.8V is used throughout, and the relative dielectric constants of high-k and low-k dielectrics are set to 20 and 1 (i.e. air gap) respectively unless mentioned otherwise. The total thickness of the device equals 4.4nm by default, including the body thickness of the channel ($\approx$0.7nm). All of the transport simulations have been performed with the simulation tool NEMO5 \cite{nemo5_2}.

\setlength{\belowdisplayskip}{1pt} \setlength{\belowdisplayshortskip}{1pt}
\setlength{\abovedisplayskip}{1pt} \setlength{\abovedisplayshortskip}{1pt}
\vspace{-0.8\baselineskip}               

\section{Simulation results}

Fig. \ref{fig:Fig2}a shows transfer characteristics of a monolayer WTe$_2$ DE-TFET with an ON-current of about 1000 uA/um and SS of 16 mV/dec. This is a record I$_{\rm ON}$-value among all reported ON-currents from full band  atomistic simulations of planar homojunction steep devices \cite{Seabaugh}. Having a high ON-current without using a hetero-junction and chemical doping has advantages in terms of OFF-state performance. Chemical doping can introduce dopant states within the band gap \cite{sapan}, and hetero-junctions often suffer from interface states. Both of these effects increase SS of the device. Fig. \ref{fig:Fig2}b shows the conduction and valence band profiles of the WTe$_2$ DE-TFET. The electric field at the tunnel junction is about 0.9 V/nm. This large value of electric field and ON-current is due to the presence of the low-k dielectric ($\epsilon_{2} \ll \epsilon_{1} $). 

The ON-current of a WTe$_2$ DE-TFET as a function of low-k spacing is shown in Fig. \ref{fig:Fig2}c. The optimum spacing $S$ to gain the maximum ON-current depends on the total thickness of the device (shown as $T_{tot}$ in Fig. \ref{fig:Fig2}a). For example, the optimum $S$ varies from about 1nm to 2nm when $T_{tot}$ changes from 4.4nm to 10.2nm. Fig. \ref{fig:Fig2}d compares the performance sensitivity of the WTe$_2$ DE-TFET with the conventional ED-TFET. The performance of the DE-TFET is much less sensitive to oxide thickness variations if compared to the conventional ED-TFET. 



\section{Analytical Modeling}
\label{sec:analytic_sec}

It is apparent that the width of the low-k dielectric ($S$) affects the performance of DE-TFETs. When $S$ equals 0, the device converts to a conventional ED-TFET and the electric field at the tunnel junction reduces. On the other hand, when $S$ becomes very large, the electric field reduces to 0. Accordingly, there exists an optimum spacing where the electric field is high. Notice that achieving the highest electric field is not a sufficient condition to ensure best performance. For example, if the electric field is high only over a very small distance such that the potential drop across the low-k dielectric is not large enough (compared to the band gap), then the tunneling window with high field will be small.

\begin{figure}[h!]
        \centering
        \begin{subfigure}[b]{0.25\textwidth}
               \includegraphics[width=\textwidth]{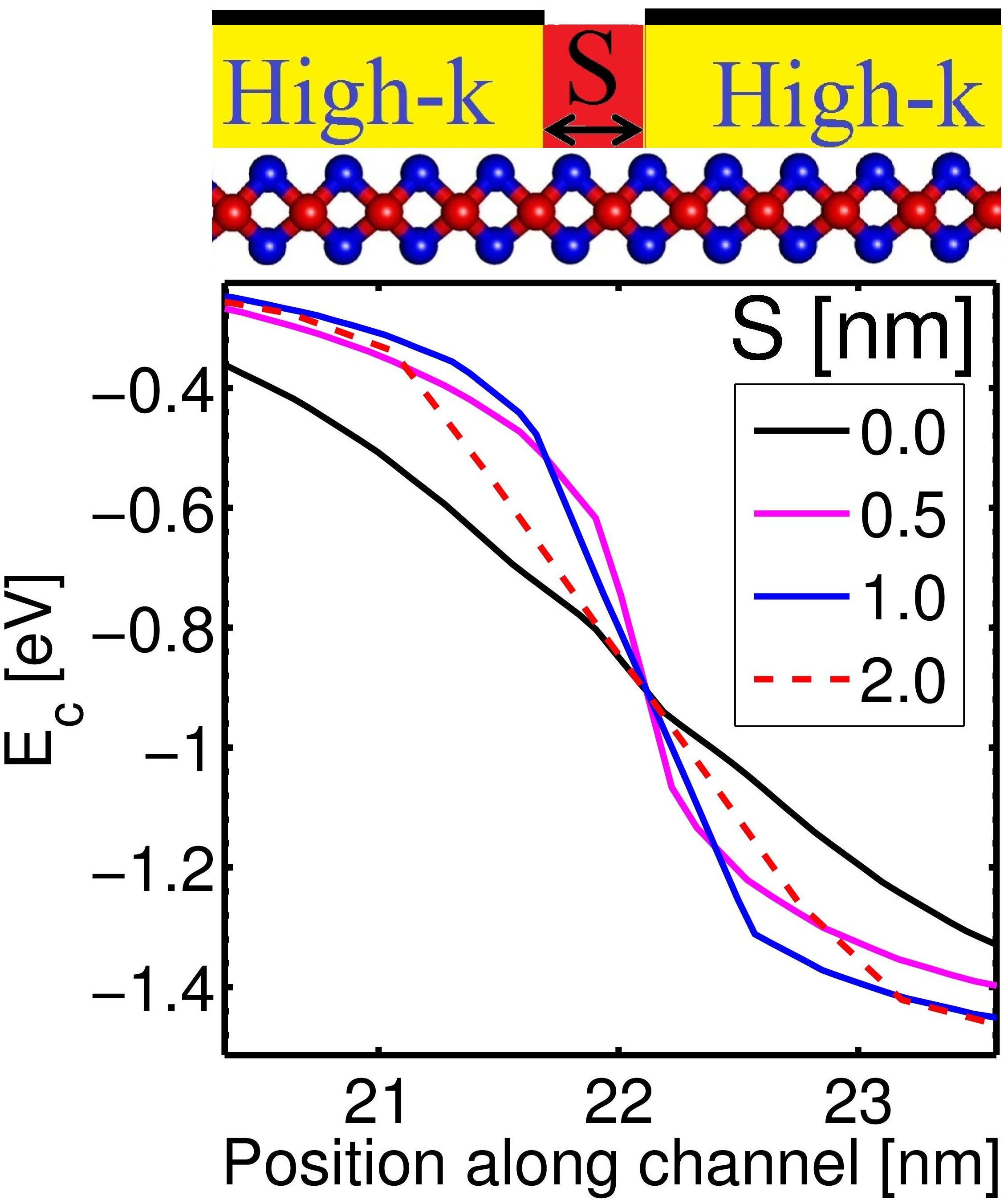}
               \vspace{-1.5\baselineskip}               
                \caption{}
                \label{fig:laplace}
        \end{subfigure}%
                \begin{subfigure}[b]{0.21\textwidth}
               \includegraphics[width=\textwidth]{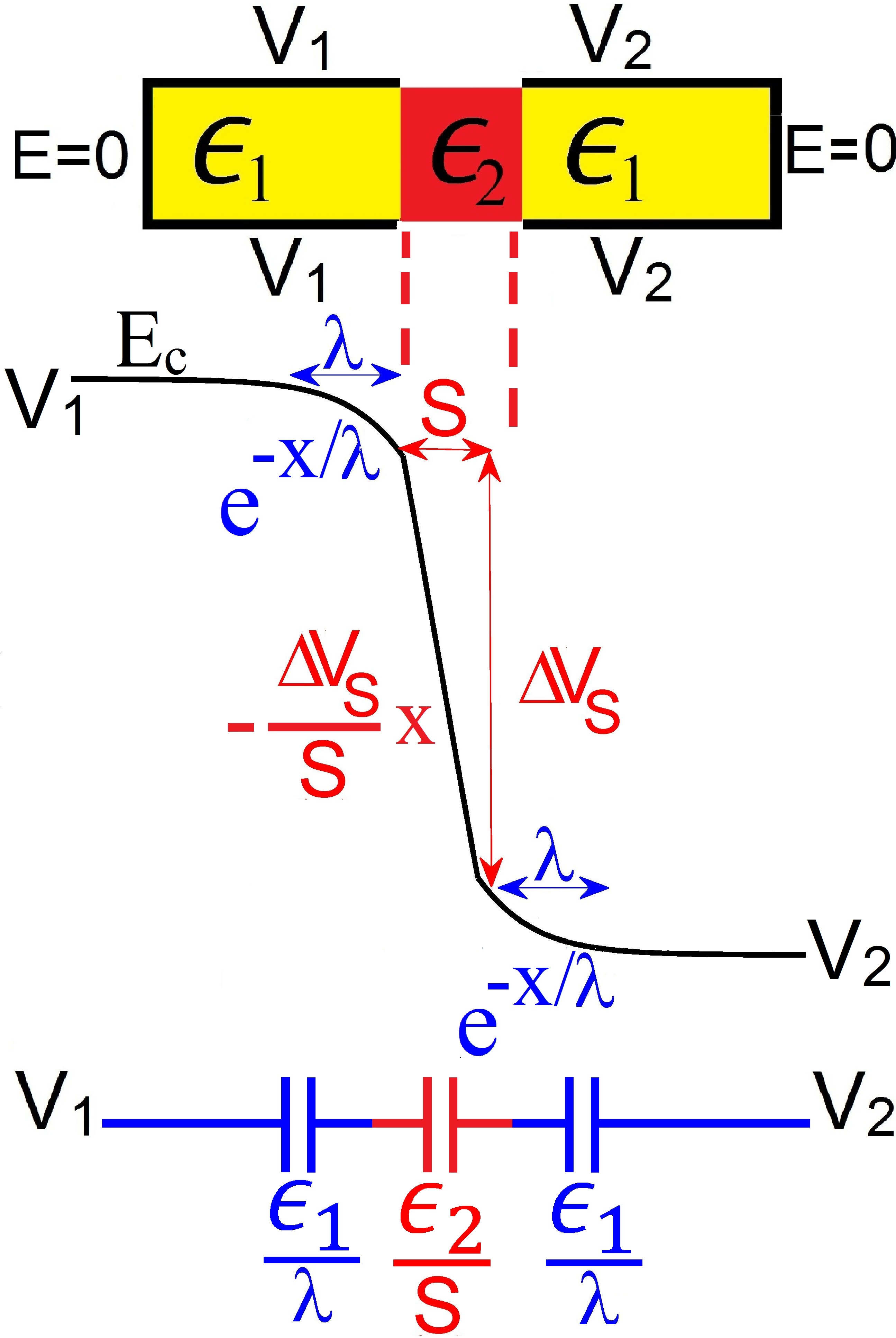}
               \vspace{-1.5\baselineskip}               
                \caption{}
                \label{fig:laplace}
        \end{subfigure}%
        \quad        
        \begin{subfigure}[b]{0.23\textwidth}
               \includegraphics[width=\textwidth]{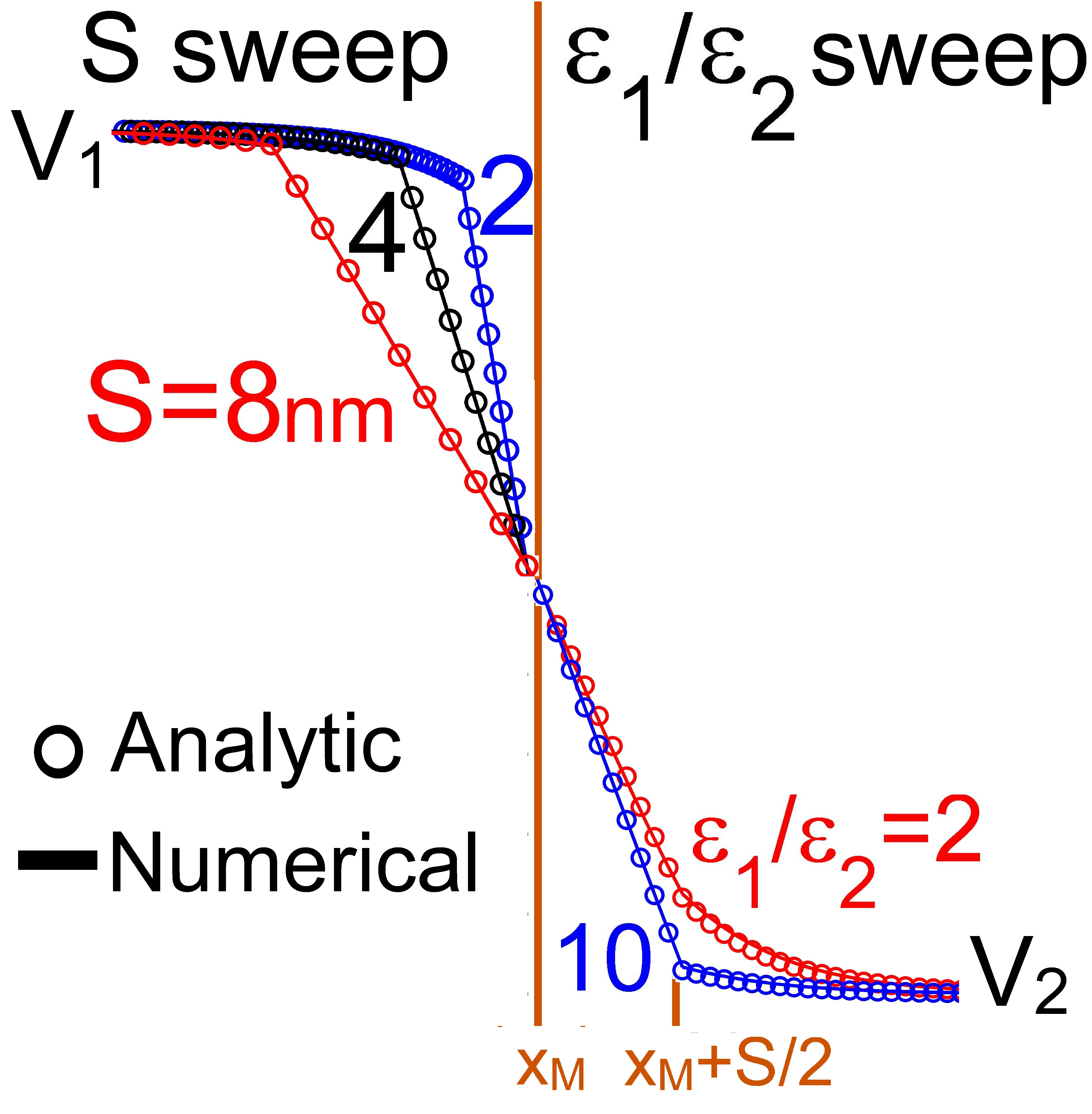}
               \vspace{-1.25\baselineskip}                              
                \caption{}
                \label{fig:st3}
        \end{subfigure}%
        \begin{subfigure}[b]{0.25\textwidth}
               \includegraphics[width=\textwidth]{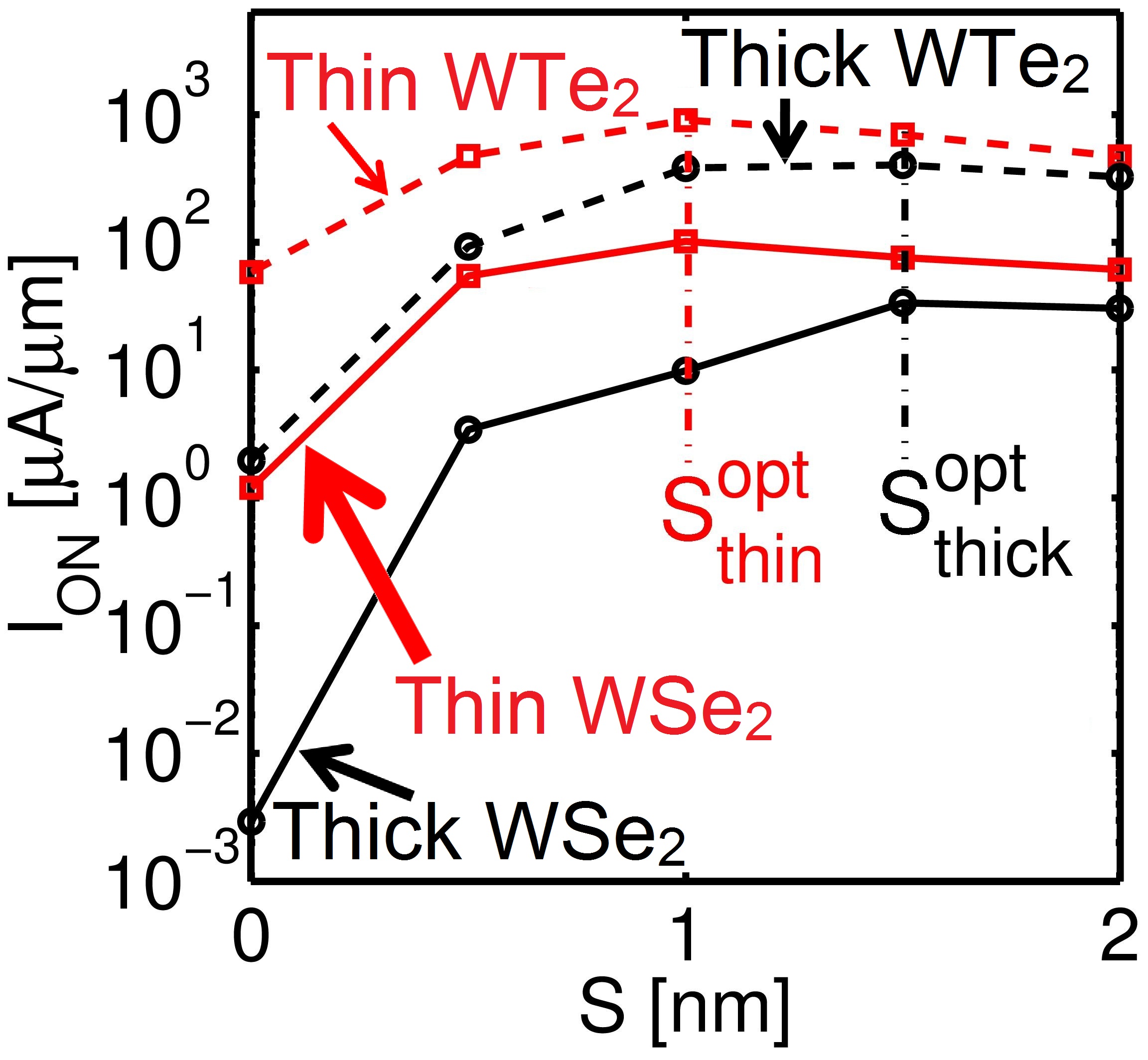}
               \vspace{-1.25\baselineskip}                              
                \caption{}
                \label{fig:st3}
        \end{subfigure}%
        \vspace{-.5\baselineskip}       
        \caption{a) Conduction band profile ($E_c$) of WTe$_2$ DE-TFETs with different $S$ obtained from atomistic simulations. b) Different sections of $E_c$ profile and equivalent capacitance network. c) Comparison between the analytic model (circle symbols) and numerical solution (solid lines) for various $S$ (left side) and $\epsilon_1 / \epsilon_2$ values (right side). d) Comparing $S^{opt}$ in WTe$_2$ (dashed lines) and WSe$_2$ (solid lines) DE-TFETs with a $T_{tot}$ of 4.4nm (red curves labeled as thin) and 7.3nm (black curves labeled as thick).}\label{fig:Fig3}	

\end{figure}

Fig. \ref{fig:Fig3}a shows the conduction band profile for a WTe$_2$ DE-TFET with different spacing values. The lower the spacing, the higher is the electric field. However, this does not always translate into better performance. For example, a spacing of 0.5nm has a larger electric field compared to the spacing of 1.0nm but the energy window with high electric field of $S=0.5nm$ case is smaller than that of $S=1.0nm$.

To understand the impact of low-k spacing $S$ on the potential profile and performance, an analytic model is developed. For the case of $S=0$, it was previously proven that the potential can be described by a $exp(-x/\lambda)$ dependence with $\lambda=\frac{T_{tot}}{\pi}$ \cite{Hesam2}. For the sake of simplicity, the channel thickness is assumed to be 0 first ($t_{ch}=0$). Later on, an empirical term is added to include the impact of $t_{ch}$. Simulations reveal that in the case of DE-TFETs, the potential is linear within the low-k dielectric as shown in Fig. \ref{fig:Fig3}b. Accordingly, the potential profile can be approximated as:
\begin{equation}
\label{eq:analytic}
\Scale[1.0]{
V\approx 
\begin{cases}
-\frac{(V_1 - V_2 - \Delta V_S)}{2} {exp}\left(+\frac{x-x_M+\frac{S}{2}}{\lambda} \right) + V_1 & x < x_M - \frac{S}{2}  \\
-\frac{\Delta V_S}{S} (x-x_M)+ \frac{(V_1 + V_2)}{2} & |x-x_M| < \frac{S}{2}\\
\frac{(V_1 - V_2 - \Delta V_S)}{2} {exp}\left(-\frac{x-x_M+\frac{S}{2}}{\lambda} \right) + V_2 & x > x_M + \frac{S}{2}  \\
\end{cases}}
\end{equation}
Here, $\Delta V_S$ is the potential drop across the low-k dielectric and $x_M$ is the position of the center of the low-k dielectric. To determine $\Delta V_S$, a capacitor network is used as shown in Fig. \ref{fig:Fig3}b. 
\begin{equation}
\label{eq:dVs}
\Delta V_S = \frac{S}{S+ 2\frac{\epsilon_2}{\epsilon_1} \lambda + \delta_{ch}} (V_1-V_2)
\end{equation}
where $\delta_{ch}$ is an empirical term to include the impact of the channel ($\delta_{ch} = 2t_{ch}/\pi$). Fig. \ref{fig:Fig3}c compares the analytic potential using (\ref{eq:analytic}) and (\ref{eq:dVs}) with the numerical solution of the 2D Poisson equation which shows that the analytic solution captures the effect of the low-k spacing $S$ and the ratio of dielectric constants ($\epsilon_1/\epsilon_2$) accurately. The maximum performance as a function of $S$ occurs when $\Delta V_S$ becomes a significant portion of the total band bending ($\frac{\Delta V_S}{V_1-V_2} \approx 0.6-0.7$). 
\begin{equation}
\label{eq:Sopt}
S^{opt} \approx t_{ch} + \frac{\epsilon_2}{\epsilon_1} T_{tot}
\end{equation}
Fig. \ref{fig:Fig3}d depicts the optimum spacing ($S^{opt}$) for the best performance of WTe$_2$ and WSe$_2$ DE-TFETs. Notice that $S^{opt}$ does not depend on the material significantly. 
\setlength{\belowdisplayskip}{2pt} \setlength{\belowdisplayshortskip}{2pt}
\setlength{\abovedisplayskip}{2pt} \setlength{\abovedisplayshortskip}{2pt}
\vspace{-0.8\baselineskip}               
\vspace{-2.0\baselineskip}       

\section{Conclusion}
In summary, a new high performance TFET is proposed with a simulated record ON-current of 1000 uA/um and SS of 16 mV/dec. The high performance and high electric field in DE-TFETs is a result of the combination of low-k and high-k dielectrics. Moreover, DE-TFETs offer other advantages: e.g. smaller sensitivity to the oxide thickness compared to conventional ED-TFETs.
The same idea can be applied to other device structures (e.g. nanowire TFETs) and other channel materials (e.g. III-V materials) to bring the ON-current of TFETs on par with MOSFETs and keep their excellent OFF-state performance intact.
\end{document}